%% file: ms.tex
\documentclass[conference]{IEEEtran}
\usepackage{ifpdf}

\usepackage{cite}

\usepackage{color}
\usepackage{soul}

\ifCLASSINFOpdf
   \usepackage[pdftex]{graphicx}
\else
   \usepackage[dvips]{graphicx}
\fi
\usepackage{amsmath}
\usepackage{algorithmic}

\usepackage{array}

\ifCLASSOPTIONcompsoc
  \usepackage[caption=false,font=normalsize,labelfont=sf,textfont=sf,subrefformat=parens,labelformat=parens]{subfig}
\else
  \usepackage[caption=false,font=footnotesize,subrefformat=parens,labelformat=parens]{subfig}
\fi

\usepackage{stfloats}
\fnbelowfloat
\usepackage{url}

\begin{document}
%
\title{Understanding Service Integration of Online\\
Social Networks: A Data-Driven Study}



%
\author{\IEEEauthorblockN{Fei Li\IEEEauthorrefmark{1}, Yang Chen\IEEEauthorrefmark{1}, Rong Xie\IEEEauthorrefmark{1}, Fehmi Ben Abdesslem\IEEEauthorrefmark{2}, Anders Lindgren\IEEEauthorrefmark{2}}
\IEEEauthorblockA{\IEEEauthorrefmark{1}School of Computer Science, Fudan University, Shanghai, China}
\IEEEauthorblockA{\IEEEauthorrefmark{2}RISE SICS, Kista, Sweden}
{Email: \{feili14, chenyang, xieronglucy\}@fudan.edu.cn, \{fehmi.ben.abdesslem, anders.lindgren\}@ri.se}
}


\maketitle

\begin{abstract}
The cross-site linking function is widely adopted by online social networks (OSNs). 
This function allows a user to link her account on one OSN to her accounts on other OSNs.  
Thus, users are able to sign in with the linked accounts, share contents among these accounts and import friends from them. It leads to the service integration of different OSNs. 
This integration not only provides convenience for users to manage accounts of different OSNs, but also introduces usefulness to OSNs that adopt the cross-site linking function. 
In this paper, we investigate this usefulness based on users' data collected from a popular OSN called Medium. 
We conduct a thorough analysis on its social graph, and find that the service integration brought by the cross-site linking function is able to change Medium's social graph structure and attract a large number of new users. However, almost none of the new users would become high PageRank users (PageRank is used to measure a user's influence in an OSN). To solve this problem, we build a machine-learning-based model to predict high PageRank users in Medium based on their Twitter data only. This model achieves a high F1-score of 0.942 and a high area under the curve (AUC) of 0.986. Based on it, we design a system to assist new OSNs to identify and attract high PageRank users from other well-established OSNs through the cross-site linking function.
\end{abstract}

\begin{IEEEkeywords}
Service Integration, Online Social Networks, Cross-site Linking, High PageRank Users, Prediction, Medium.
\end{IEEEkeywords}

%
\IEEEpeerreviewmaketitle

\input{introduction}
\input{related_work}
\input{data_collection}

\input{data_analysis}

\input{conclusion}

\section*{Acknowledgement}

This work is sponsored by National Natural Science Foundation of China (No. 61602122, No. 71731004), Natural Science Foundation of Shanghai (No. 16ZR1402200), Shanghai Pujiang Program (No. 16PJ1400700). Yang Chen is the corresponding author.

\bibliographystyle{IEEEtran}
\bibliography{ms}



%
%


\end{document}

%% file: introduction.tex
\section{Introduction}
\label{Introduction}
Numerous online social networks (OSNs) have emerged in recent years. Many of them, such as Foursquare, Pinterest, Quora and Medium, have enabled the cross-site linking function~\cite{Chen:2014:UCL:2659480.2659498}. 
This function allows a user to link her account on one OSN to her accounts on other OSNs. 
By linking to other accounts, users are able to sign in with the linked accounts, share contents to other OSNs and import friends from them. It leads to the service integration between different OSNs. This integration provides convenience for users to manage accounts of different OSNs. More importantly, it can influence the OSNs who adopt the cross-site linking function.

Our goal is to better understand the usefulness of this service integration on OSNs. 
We introduce a data-driven study to investigate it. 
We choose the popular online publishing platform Medium as our main research focus. It has enabled the cross-site linking to Twitter and Facebook. 
It has been possible to link Twitter accounts to Medium since the launch of Medium in August 2012, and 2 years later Facebook was able to be linked to it. In 2016, Medium grew 140\% to 60 million monthly visitors, largely due to the turbulent political climate as articles about Brexit and Trump contributed a lot to its popularity.~\footnote{https://venturebeat.com/2016/12/14/medium-grows-140-to-60-million-monthly-visitors}

In this work, we have collected data about more than 1 million Medium users, covering about 25\% of the total users in August 2016.~\footnote{We estimate the number of Medium users by looking at the official account ``Medium Staff'', which is automatically followed by each user after registration. ``Medium Staff'' had 3.8 million followers in August 2016, whereas it is followed by more than 11.5 million users now (November 2017), so there were around 4 million users in total at our collection time.} We construct a social graph of the users of Medium based on the collected data and conduct a basic analysis on it. Next, we study the usefulness of service integration on Medium by analyzing the dynamics of the Medium social graph and the behavioral difference among Medium users having different cross-site linking options, i.e., whether they have linked their Medium account to Twitter and/or Facebook accounts. We discover that despite Medium's enabling of the cross-site linking function to Facebook can attract a large number of new users, almost none of them would become high PageRank users (PageRank is used to measure a user's influence in an OSN~\cite{Kwak:2010:TSN:1772690.1772751,Tang:2012:IST:2124295.2124382,Liu:2017:IPV:3058790.3046941}). 
To assist new OSNs to attract more high PageRank users through the cross-site linking function, we build a machine-learning-based classification model and extend it to a system.

Our main contributions are summarized below.
\begin{itemize}
  \item We analyze the Medium social graph to obtain a general view of this OSN. To the best of our knowledge, it is the first analysis of Medium's social graph.
  \item We study the usefulness of service integration on OSNs by analyzing the dynamics of the Medium social graph. We find that Medium's enabling of the cross-site linking to Facebook brings a change of graph structure and a large number of new users to it. However, almost none of them would become high PageRank users in Medium. 
  \item We build a machine-learning-based classification model to assist new OSNs like Medium to identify potential high PageRank users from established OSNs like Twitter and Facebook. It predicts high PageRank Medium users based on Twitter data only and achieves an F1-score of 0.942 and an area under the curve (AUC) of 0.986. We also design a system based on our model to assist new OSNs to attract high PageRank users from well-established OSNs through the cross-site linking function.
\end{itemize}

The rest of this paper is organized as follows. We review the related work in Section~\ref{related_section}, and introduce the Medium data set in Section~\ref{data_collection_section}. Then, to understand the usefulness of service integration on OSNs, we analyze the collected data in Section~\ref{data_analysis_section}. Finally, we conclude the work in Section~\ref{conclude_section}.

%% file: related_work.tex
\section{Related Work}
\label{related_section}

Some previous works studied the integration of different social networks. Zhao \textit{et al.}~\cite{Zhao:2012:MDM:2398776.2398795} studied a rare online social network (OSN) merge event, i.e. an OSN merged with its largest competitor. In our work, the service integration of OSNs is not a network-level merge like that. It is caused by the sharing of user behaviors and activities among different OSNs through the cross-site linking function.

There are also works that focus on the cross-site linking function. Zhong \textit{et al.}~\cite{Zhong:2014:SBP:2566486.2568031} studied the function in Pinterest and Last.fm focusing on how social bootstrapping from established OSNs can help engage new users to new OSNs. Chen \textit{et al.}~\cite{Chen:2014:UCL:2659480.2659498} found that Foursquare users who have enabled the cross-site linking function are more active than other users. In our work, despite the benefits, we find a shortcoming of Medium's enabling of a new cross-site linking option. Although the option brings a large number of new users from Facebook to Medium, almost none of them become high PageRank users.

Prediction models based on aggregated data of different OSNs have also been studied. Liu \textit{et al.}~\cite{Liu:2014:HLS:2588555.2588559} built a model to identify user linkage across different OSNs to obtain more accurate profiling of users. Goga \textit{et al.}~\cite{Goga:2013:EIA:2488388.2488428} built a model to detect accounts belonging to the same user in different OSNs with a focus of user privacy. Differently, we build a model to predict high PageRank users of one OSN based on data of the other OSN only. We design a system based on the model to assist new OSNs to attract high PageRank users from other well-established OSNs through the cross-site linking function.

\begin{table*}[!t]
\renewcommand{\arraystretch}{1.3}
\caption{Analysis of the Medium Social Graph}
\label{graph_table}
\centering
\begin{tabular}{c | c | c}
\hline
Attribute & Definition & Value \\
\hline
Nodes & Number of nodes & 1,075,983 \\
Edges & Number of edges & 30,026,896 \\
Zero InDeg Nodes & Number of nodes with zero in-degree & 205,734 \\
Zero OutDeg Nodes & Number of nodes with zero out-degree & 1,407 \\
Users Following ``Medium Staff'' & Number of users who are following the official account ``Medium Staff'' & 1,051,242 \\
Users Only Following ``Medium Staff'' & Number of users who are only following the official account ``Medium Staff'' & 54,002 \\
Average Clustering Coefficient & Average of all nodes' local clustering coefficient & 0.36 \\
Size of LSCC & Number of nodes in the largest strongly connected components & 838,021 \\
Average Path Length & Average length of shortest paths of all nodes pairs in LSCC & 4.36 \\
\hline
\end{tabular}
\end{table*}

%% file: data_collection.tex
\section{Data Collection}
\label{data_collection_section}

From August 14th to 29th in 2016, we collected data of 1.07 million Medium users, which covered about 25\%
of the total users at that time. 
We collect users' data by crawling their profile pages which could be accessed by their user names. 
We use Breadth First Search to collect user names through following relationships between users. For each user, we scraped her following and follower lists, registration timestamp, and cross-site linking option, i.e. whether she has linked her Medium account to Twitter and/or Facebook accounts. We also collect her Twitter profile data if she has linked Medium to her Twitter account. To speed up the crawling process, 
we developed a distributed web crawler using the crowd crawling framework~\cite{Ding_Crawler_COSN13} and deployed it on 10 virtual instances of Amazon Web Services.

%% file: data_analysis.tex
\section{Data Analysis}
\label{data_analysis_section}

\begin{figure}
\centering

\subfloat[CCDF of Out-degree]{\includegraphics[width=.45\linewidth]{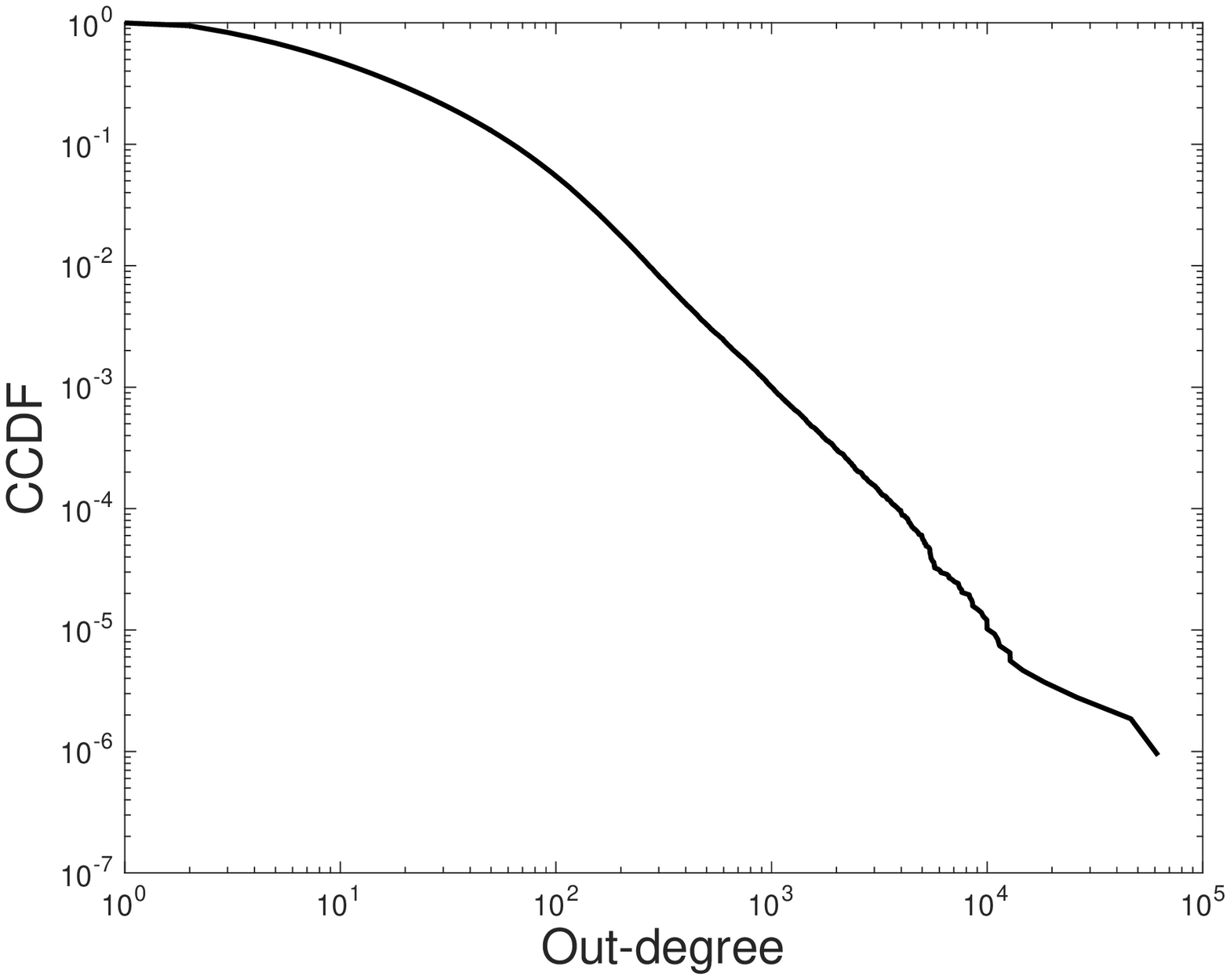}
\label{ccdf_out_degree}}
\hfill
\subfloat[CCDF of In-degree]{\includegraphics[width=.45\linewidth]{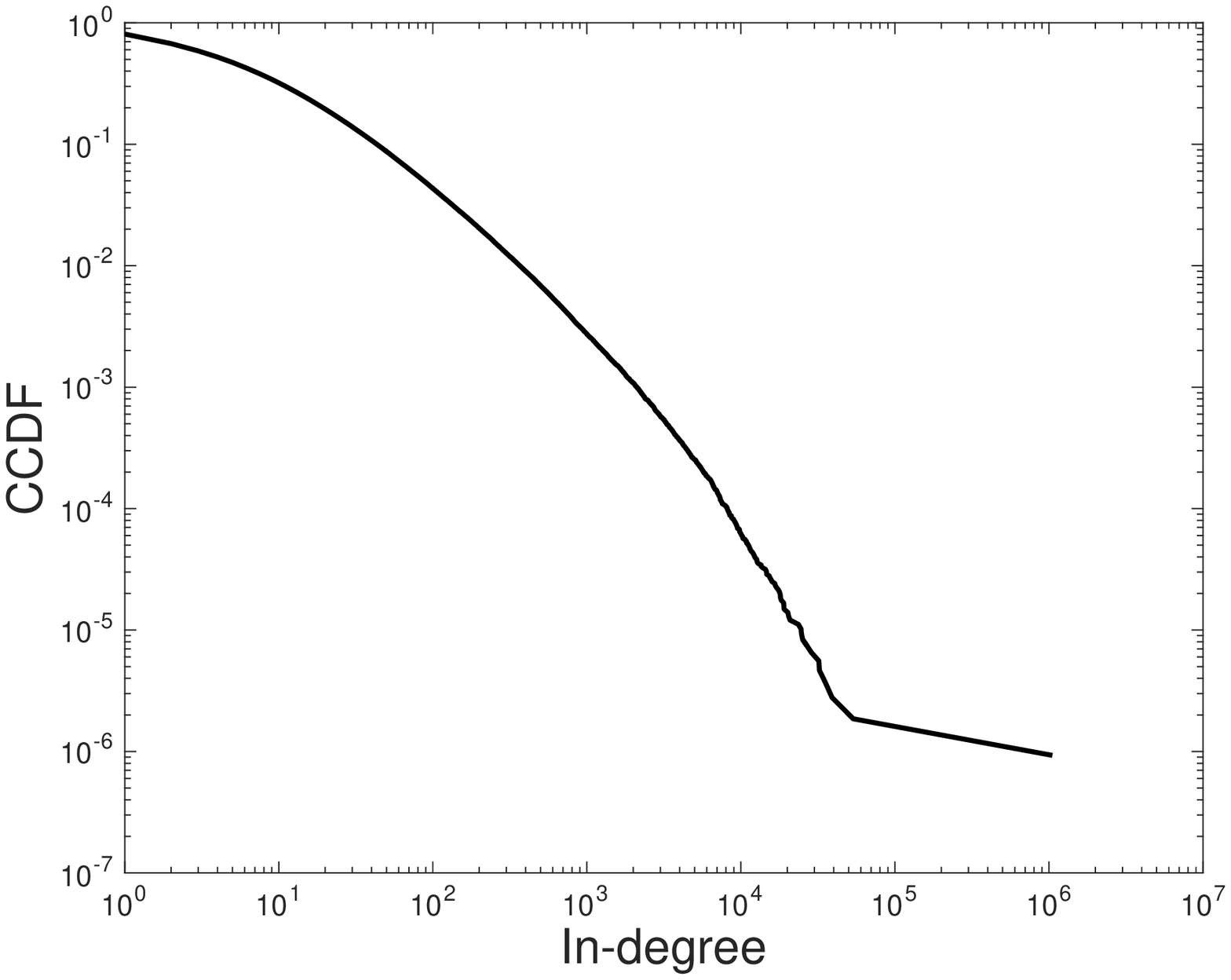}
\label{ccdf_in_degree}}

\caption{Analysis of the Medium Social Graph}
\label{ccdf_degree}
\end{figure}

In this section, we study the usefulness of service integration between different OSNs based on analysis of the data set described in Section~\ref{data_collection_section}. 
First, 
we present the analysis of the Medium social graph. 
Then, 
we study the usefulness of OSN service integration by analyzing the dynamics of the Medium social graph as well as the behavioral difference among Medium users having different cross-site linking options. 
After that, we propose a classification model to predict high PageRank Medium users based on their Twitter data only. 

\subsection{Analysis of the Medium Social Graph}
\label{measure_subsection}

Based on users' following and follower lists in the data set, we construct a large social graph of Medium as a directed graph $G=(V,E)$. A node in $V$ represents a user, and an edge in $E$ represents a following relationship. For any two nodes (users) $v_1 \in V$ and $v_2 \in V$, an edge (connection) $e \in E$ from $v_1$ to $v_2$ indicates that $v_1$ is following $v_2$ in Medium. The out-degree/in-degree of a node in $G$ indicates the number of followings/followers of the corresponding user in this social graph. The Medium social graph is a weakly connected graph since we use Breadth First Search as the crawling method.

We analyze the Medium social graph by using Stanford Network Analysis Project (SNAP) library~\cite{leskovec2016snap} as shown in Table~\ref{graph_table}.
According to the complementary cumulative distribution function (CCDF) of out-degrees and in-degrees in Fig.~\ref{ccdf_degree}, they both resemble the power-law distribution, which is similar to other OSNs (Twitter~\cite{Myers:2014:INS:2567948.2576939,Kwak:2010:TSN:1772690.1772751}, Facebook~\cite{DBLP:journals/corr/abs-1111-4503} and Google+~\cite{Gonzalez:2013:GGD:2488388.2488431}). The long tail in Fig.~\subref*{ccdf_in_degree} represents the huge number of followers of the official account ``Medium Staff''. In Table~\ref{graph_table}, we can see that 205,734 (19.12\%) users have zero follower while 1,407 (0.13\%) users have zero following, which implies that a large fraction of Medium users are not very active or has just joined to follow someone else's writings. 
Among all users, 1,051,242 (97.70\%) of them are following ``Medium Staff'', which proves our estimate of the number of total users in Section~\ref{Introduction} is reasonable. There are also 54,002 (5.02\%) users who are only following ``Medium Staff'', which indicates that these users signed up to Medium just for a trial. 
Clustering coefficient is a measure of the extent to which nodes in a graph tend to cluster together. The average clustering coefficient in Medium is 0.36, which is much higher than that in other OSNs (0.072 in Twitter~\cite{6785885}, 0.13 in Facebook~\cite{Viswanath:2009:EUI:1592665.1592675}, 0.14 in Renren~\cite{Zhao:2012:MDM:2398776.2398795}). It indicates that Medium users are densely connected. 
The largest strongly connected component covers 77.88\% of users in the Medium social graph. In the component, the average shortest path length is 4.36, which is similar to other OSNs (4.05 in Twitter~\cite{Myers:2014:INS:2567948.2576939}, 4.74 in Facebook~\cite{DBLP:journals/corr/abs-1111-4503}).  It shows the property of small-world networks~\cite{small-world} since it has both a high average clustering coefficient and a small average shortest path length.

\subsection{Analysis of the Usefulness of Service Integration}
\label{csl_subsection}

\begin{figure}
\centering

\subfloat[Dynamic Distribution of Different Linking Options]{\includegraphics[width=.45\linewidth]{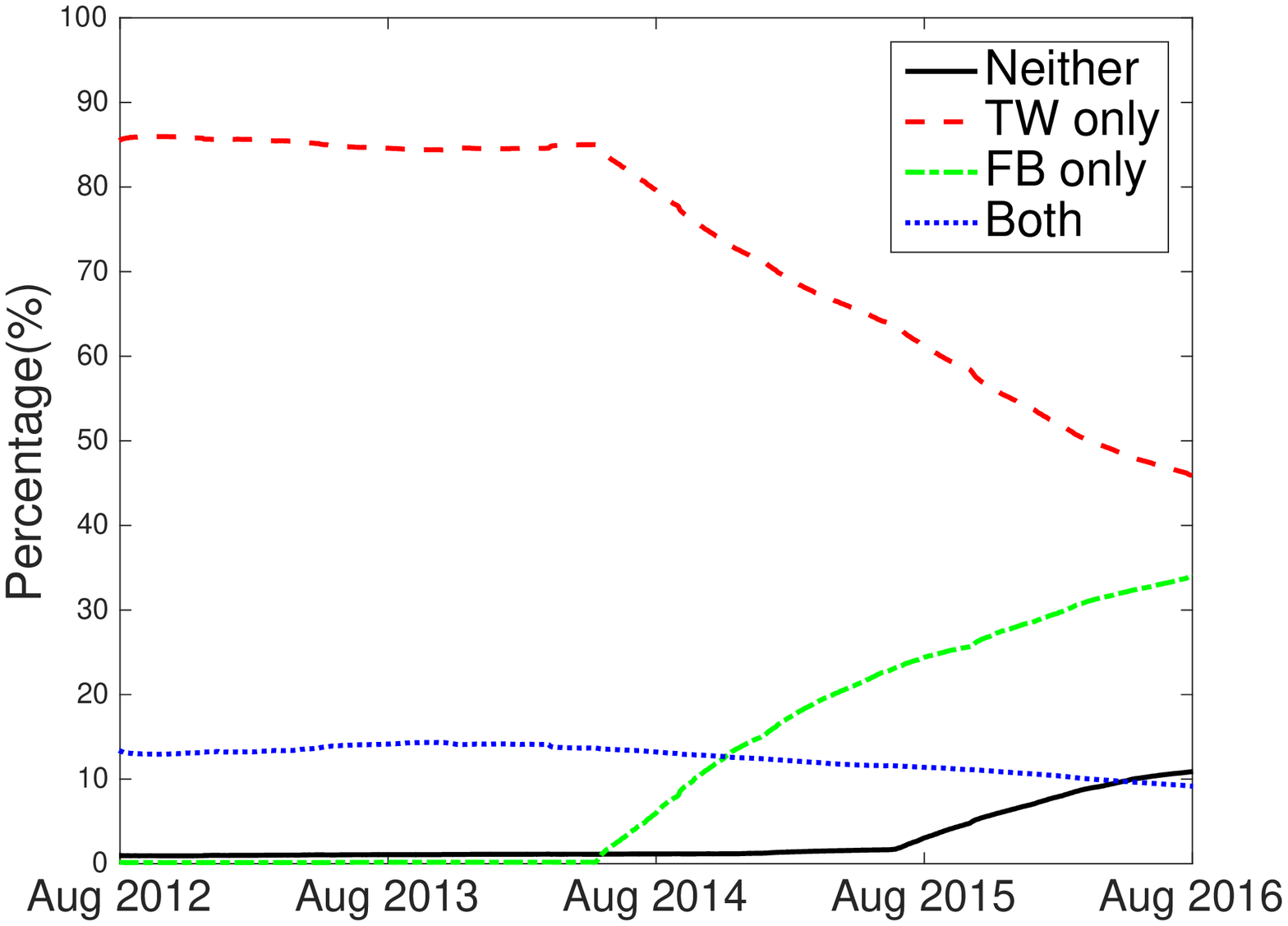}
\label{csl_time}}
\hfill
\subfloat[Dynamic Average Degree]{\includegraphics[width=.45\linewidth]{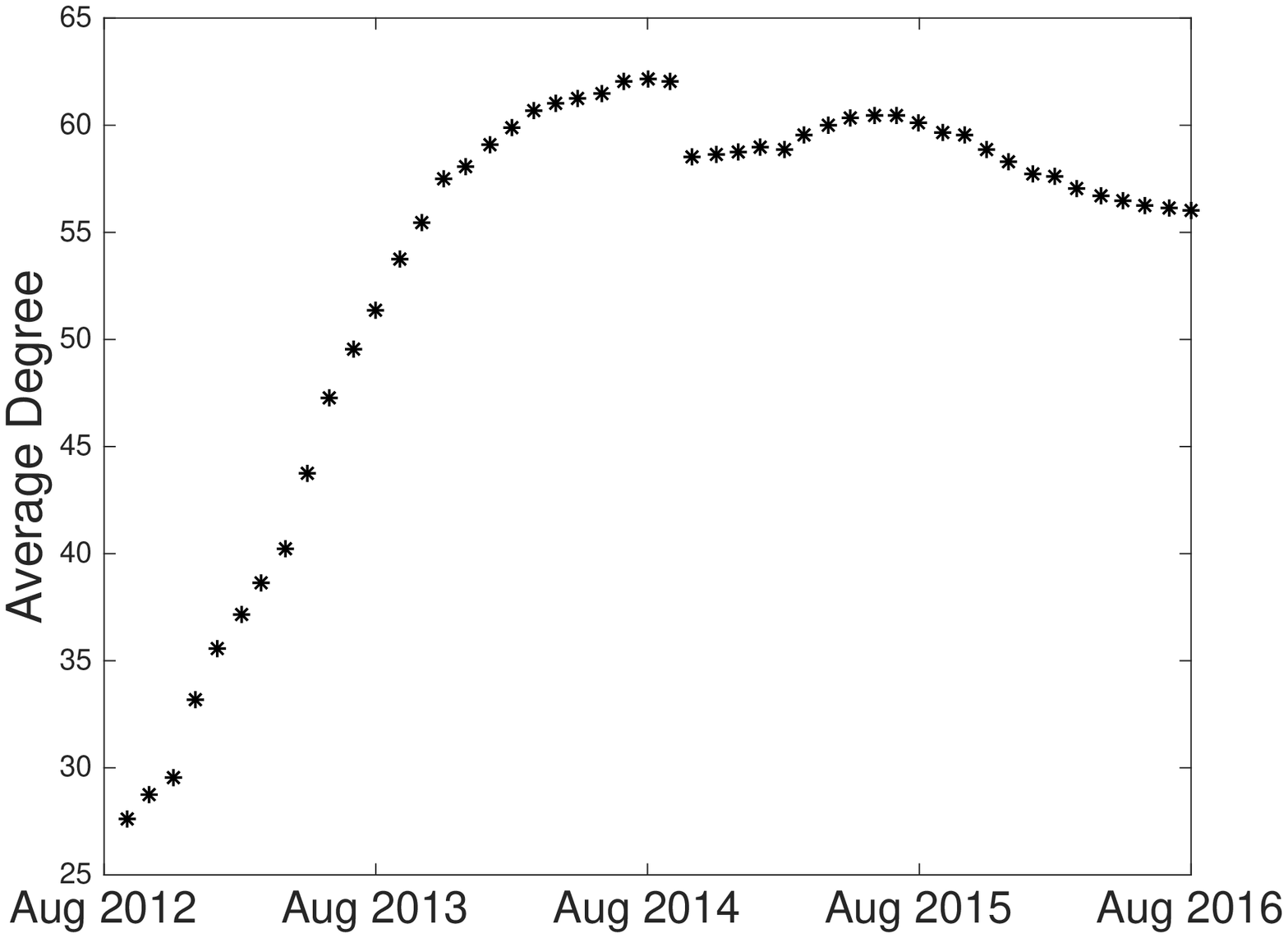}
\label{degree_time}}

\subfloat[CDF of Out-degree of Different Linking Options]{\includegraphics[width=.45\linewidth]{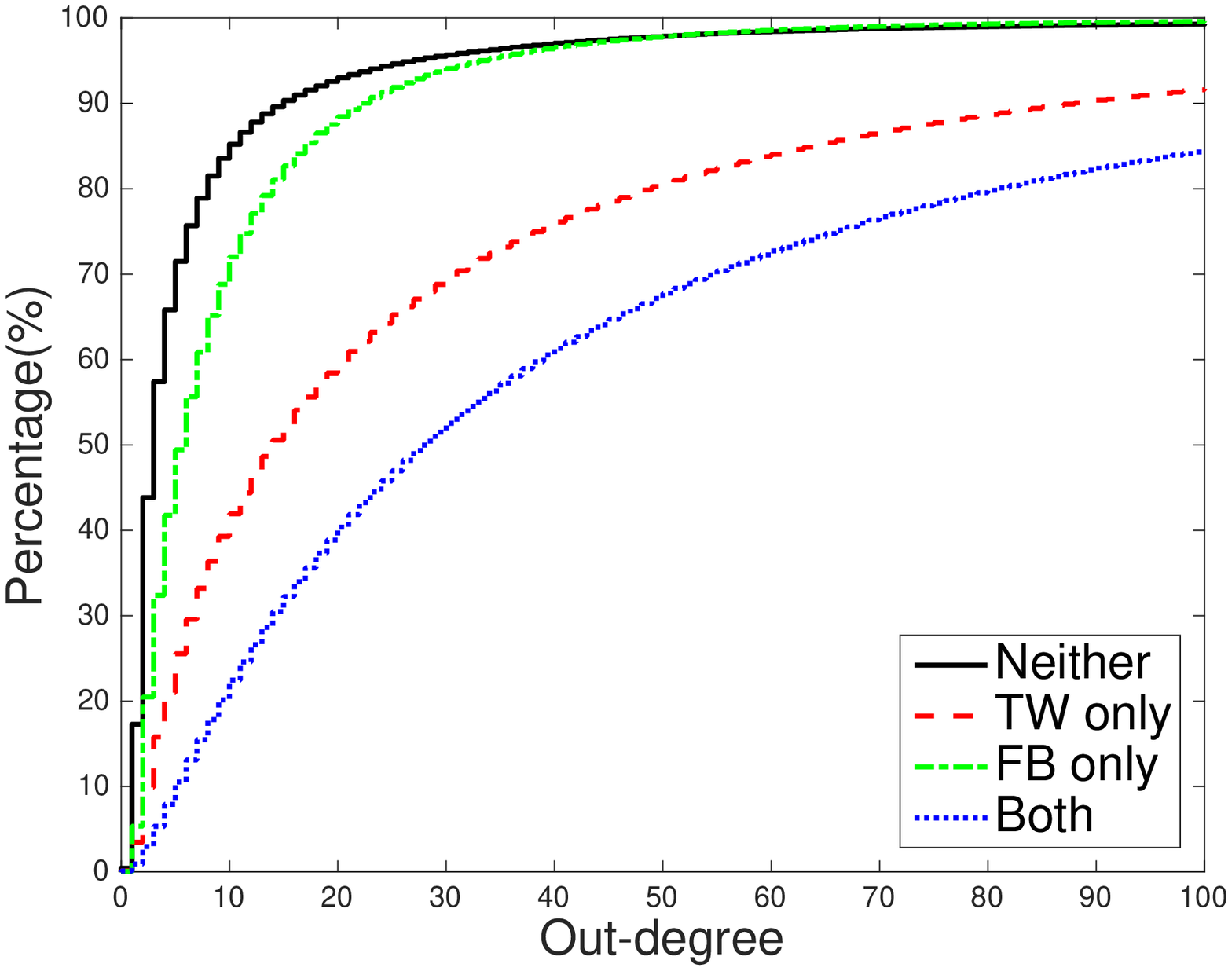}
\label{out_degree_csl}}
\hfill
\subfloat[CDF of In-degree of Different Linking Options]{\includegraphics[width=.45\linewidth]{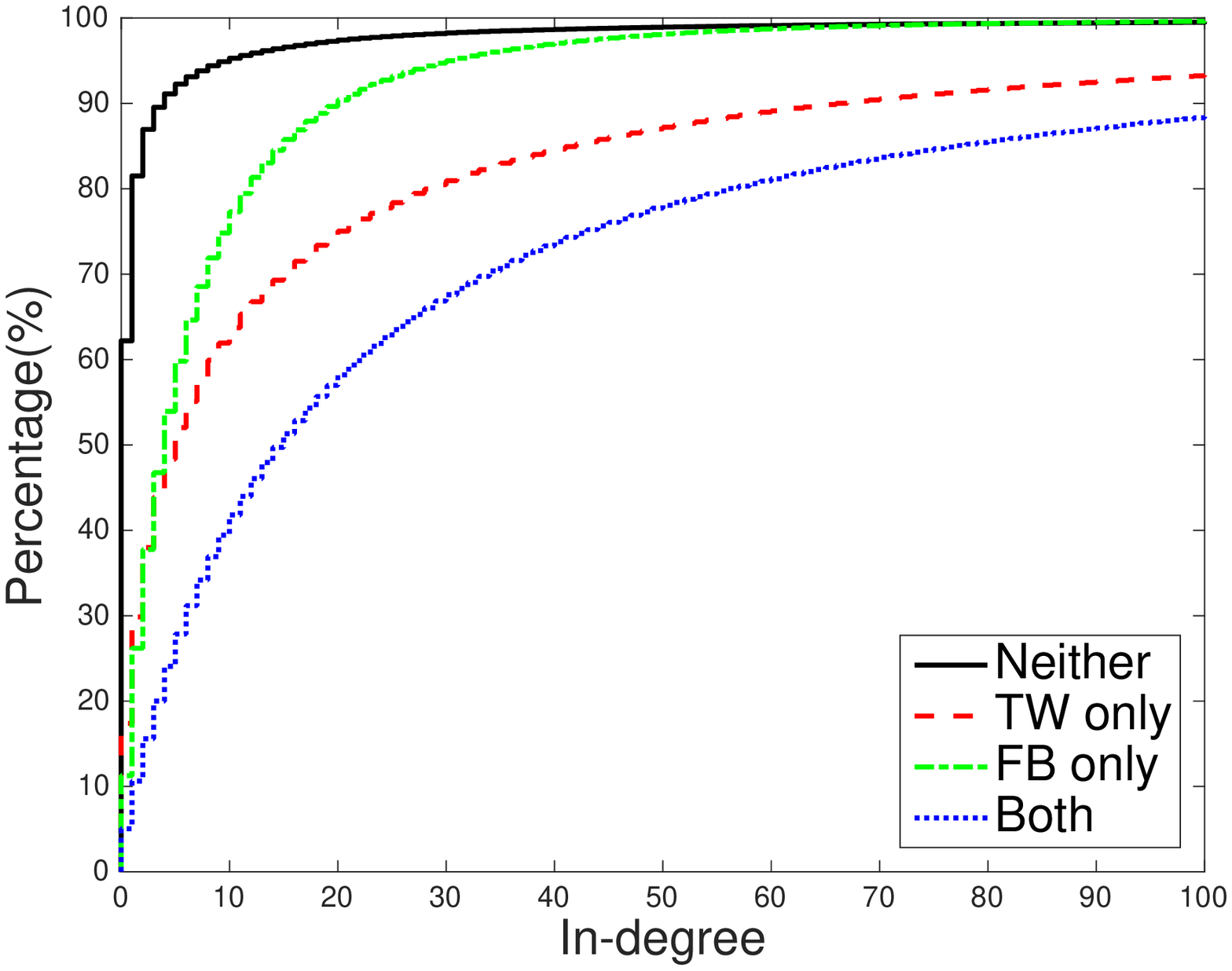}
\label{in_degree_csl}}

\subfloat[CDF of PageRank of Different Linking Options]{\includegraphics[width=.45\linewidth]{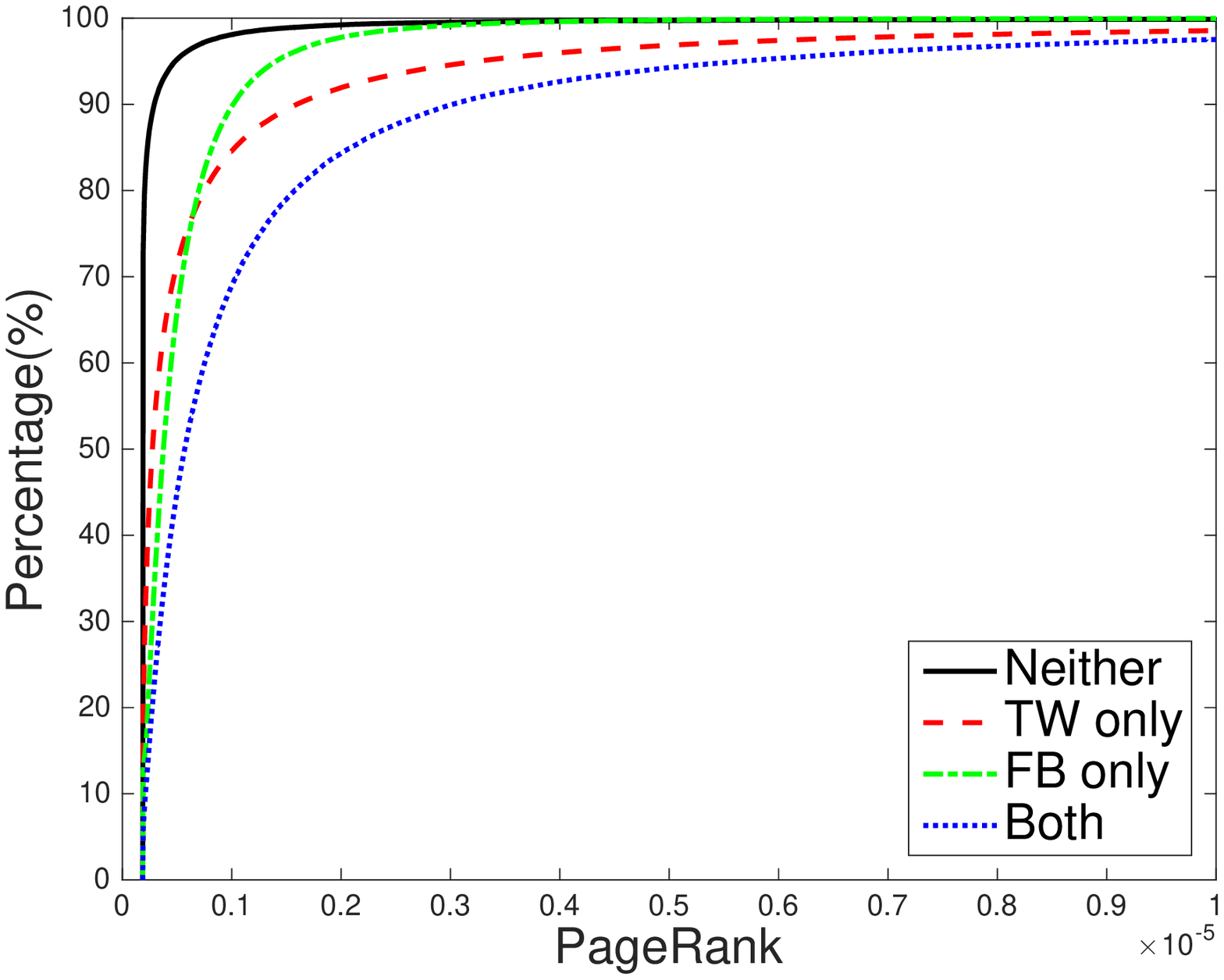}
\label{pagerank_csl}}
\hfill
\subfloat[Distribution of High and Low PageRank Users]{\includegraphics[width=.45\linewidth]{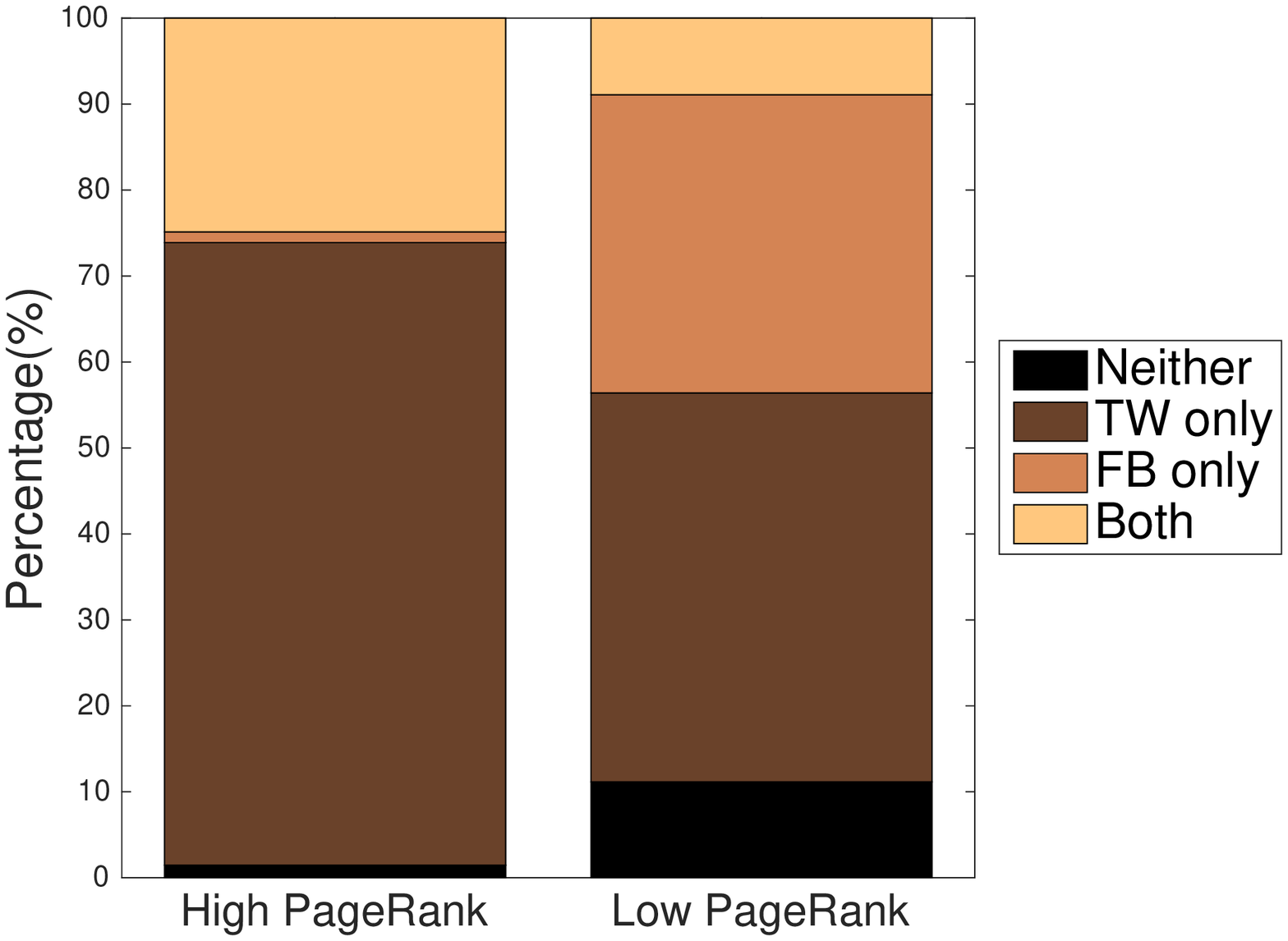}
\label{impact_csl}}

\caption{Analysis of the Usefulness of Service Integration}
\label{dynamics}
\end{figure}

In this subsection, we study the usefulness of service integration on Medium by analyzing the dynamics of its social graph and behavioral difference among its users having different cross-site linking options. 
Users can link their Medium accounts to Twitter and/or Facebook accounts. 
We assign a ``linking option'' proposed by Chen \textit{et al.}~\cite{Chen:2014:UCL:2659480.2659498} to each Medium user according to their linked accounts. There are four linking options, i.e., ``Neither'', ``TW only'', ``FB only'' and ``Both''. ``Neither'' means that the user has not linked her Medium account to any other accounts. ``TW only'' and ``FB only'' represent that the user links her Medium account to only Twitter or only Facebook account, respectively. ``Both'' means that her Medium account has been linked to both Twitter and Facebook accounts. 
11.06\%, 45.52\%, 34.35\% and 9.07\% of Medium users are assigned with ``Neither'', ``TW only'', ``FB only'' and ``Both'', respectively. These percentages imply that the cross-site linking function is widely used by Medium users. 

However, the distribution of the four linking options changes over time due to the introduction of new linking options. 
By utilizing the collected registration timestamps of each user, we generate the dynamic distribution of different linking options in Fig.~\subref*{csl_time}. It shows the change of linking option distribution from August 2012 to August 2016. We can see that on June 2014, ``FB only'' users began to grow from nearly zero, which implies that Medium enabled the Facebook linking option around that time. We can also see that some ``Both'' users have already registered before June 2014, which indicates that some previous ``TW only'' users connected to their Facebook accounts after the enabling of the ``FB'' option.

To investigate the usefulness of the service integration of Medium and Facebook, we analyze the dynamics of the Medium social graph. Since we cannot get access to the past social graphs of Medium, we use an approach proposed and validated by Gabielkov~\textit{et al.}~\cite{Gabielkov:2014:SSN:2591971.2591985} to construct past social graphs.
For each month between September 2012 and August 2016, we remove from our data set all users (nodes) registered after that month, and all following relationships (edges) to and from these users (nodes). Then we use the remaining data set to construct the Medium social graph for that month.

We conduct analysis on these past social graphs. Fig.~\subref*{degree_time} shows the dynamic average degree of the Medium social graph. We can see a significant drop of average degree during September 2014. According to Fig.~\subref*{csl_time}, we conclude that it is caused by a swarm of new users that created Medium accounts because of the new Facebook linking option. Fig.~\subref*{out_degree_csl} and Fig.~\subref*{in_degree_csl} show the CDF of out-degree and in-degree number of users having different linking options, respectively. We can see that ``FB only'' users have lower out-/in-degrees than ``TW only users'', which explains the drop of average degree.

In our further analysis of behavioral differences among users having different linking options, we find that ``FB only'' users have less influence than ``TW only'' users. As in~\cite{Kwak:2010:TSN:1772690.1772751,Tang:2012:IST:2124295.2124382,Liu:2017:IPV:3058790.3046941}, PageRank is used to measure a user's influence in an OSN. Fig.~\subref*{pagerank_csl} shows the CDF of PageRank of Medium users having different linking options. We can see that few ``FB only'' users have high PageRank. Based on Medium users' PageRank values, we can classify all Medium users into two groups, i.e., ``high PageRank users'' and ``low PageRank users''. We find 1\% of Medium users have PageRank values larger than $9.1746 \times 10^{-6}$. Thus, we consider a user as a ``high PageRank user'' if she has a PageRank value over $9.1746 \times 10^{-6}$. Otherwise, the user is marked as a ``low PageRank user''. Then we analyze the distribution of users having different linking options in the two groups. Fig.~\subref*{impact_csl} shows that only 1.25\% of high PageRank Medium users are ``FB only'' users, while ``TW only'' users cover 72.45\% of high PageRank users in Medium, indicating that almost none of the high PageRank users are ``FB only'' users.

In summary, the usefulness of service integration on Medium can be demonstrated from two perspectives. First, the introduction of new linking option can change the structure of the Medium social graph. Second, it can attract new users to Medium from Facebook. However, almost none of the new users would become high PageRank users in Medium, which is a shortcoming of the service integration.

\begin{figure}
\centering
\subfloat[Followings]{\includegraphics[width=.45\linewidth]{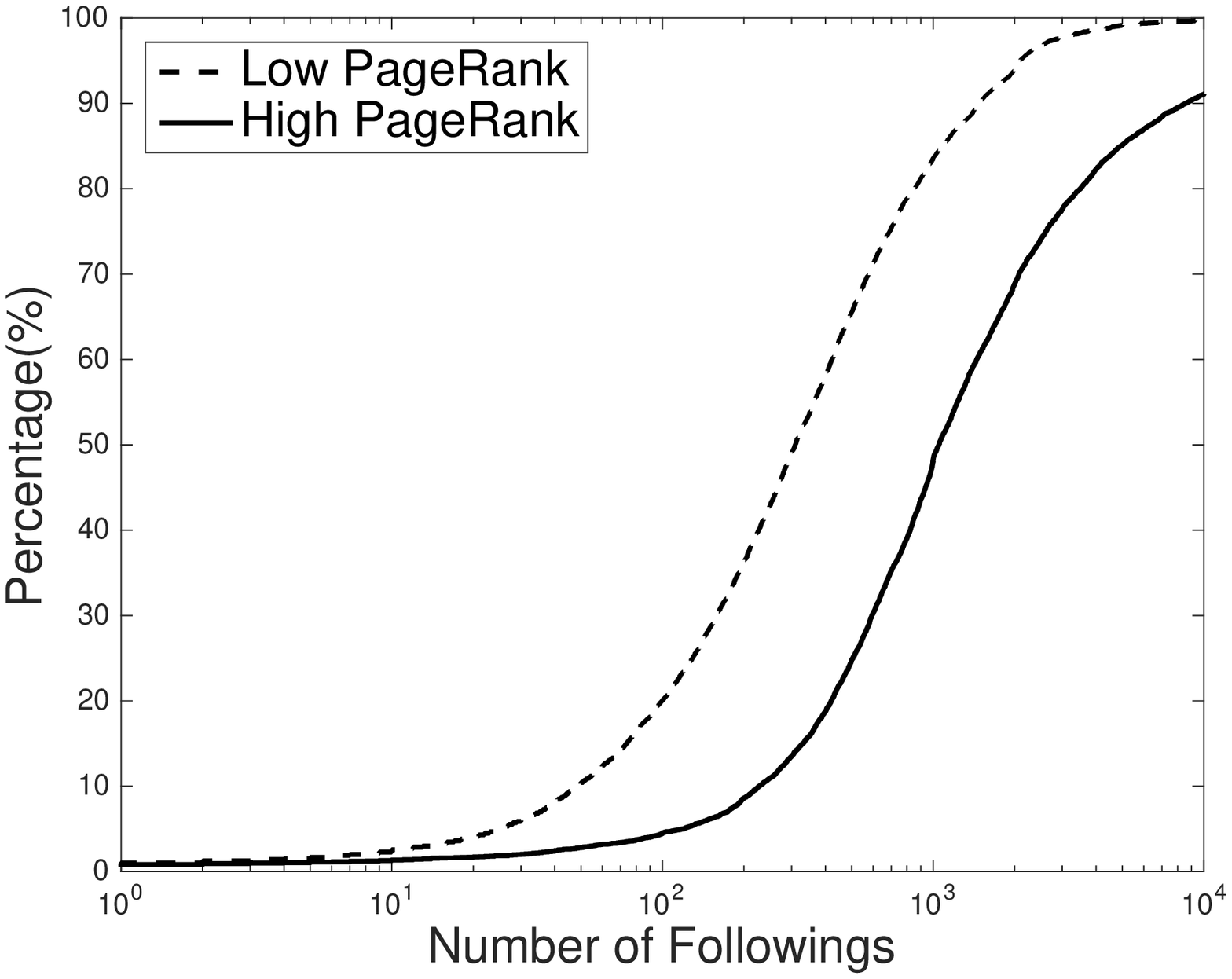}
\label{t_following}}
\hfill
\subfloat[Followers]{\includegraphics[width=.45\linewidth]{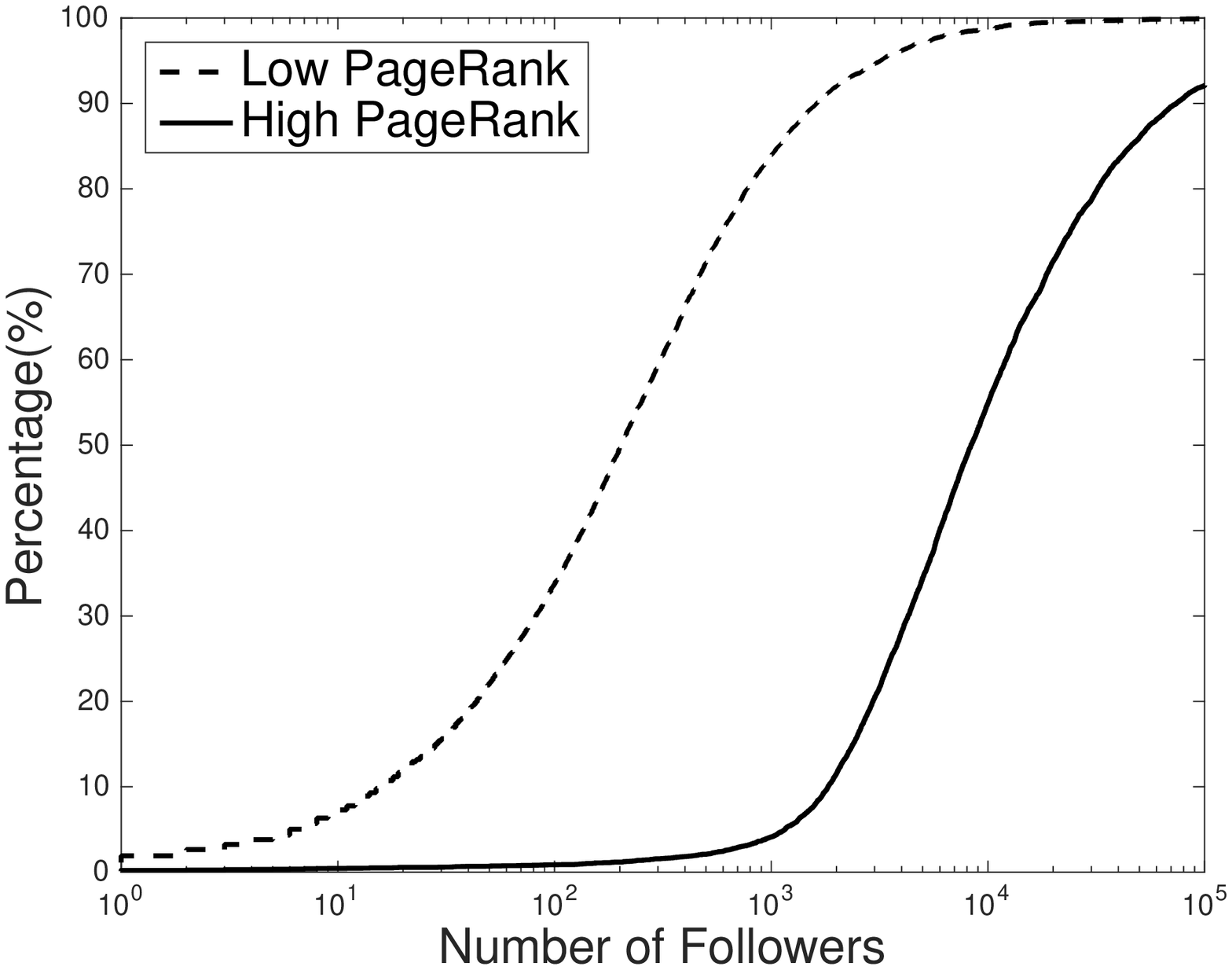}
\label{t_followers}}

\subfloat[Tweets]{\includegraphics[width=.45\linewidth]{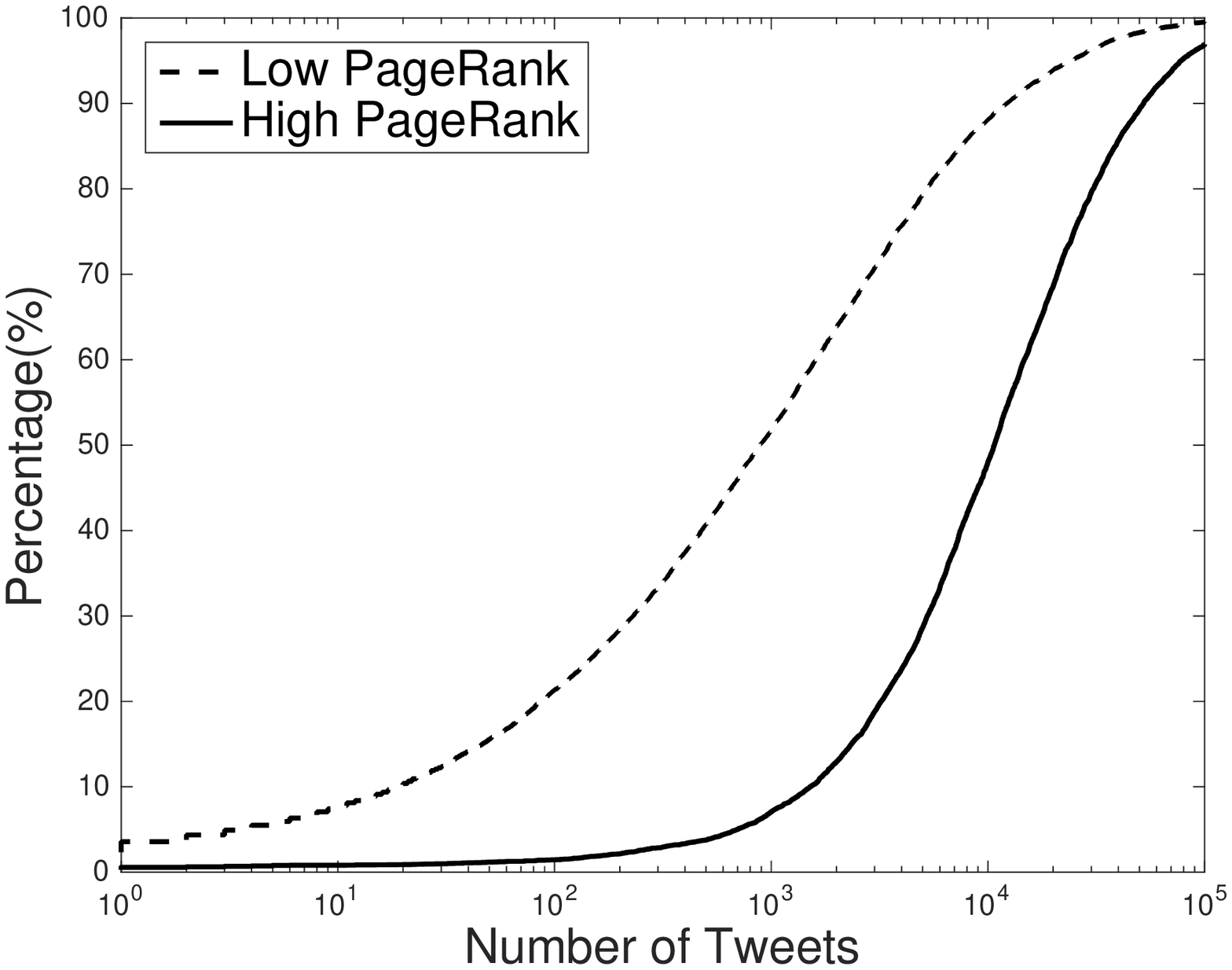}
\label{t_tweets}}
\hfill
\subfloat[Likes]{\includegraphics[width=.45\linewidth]{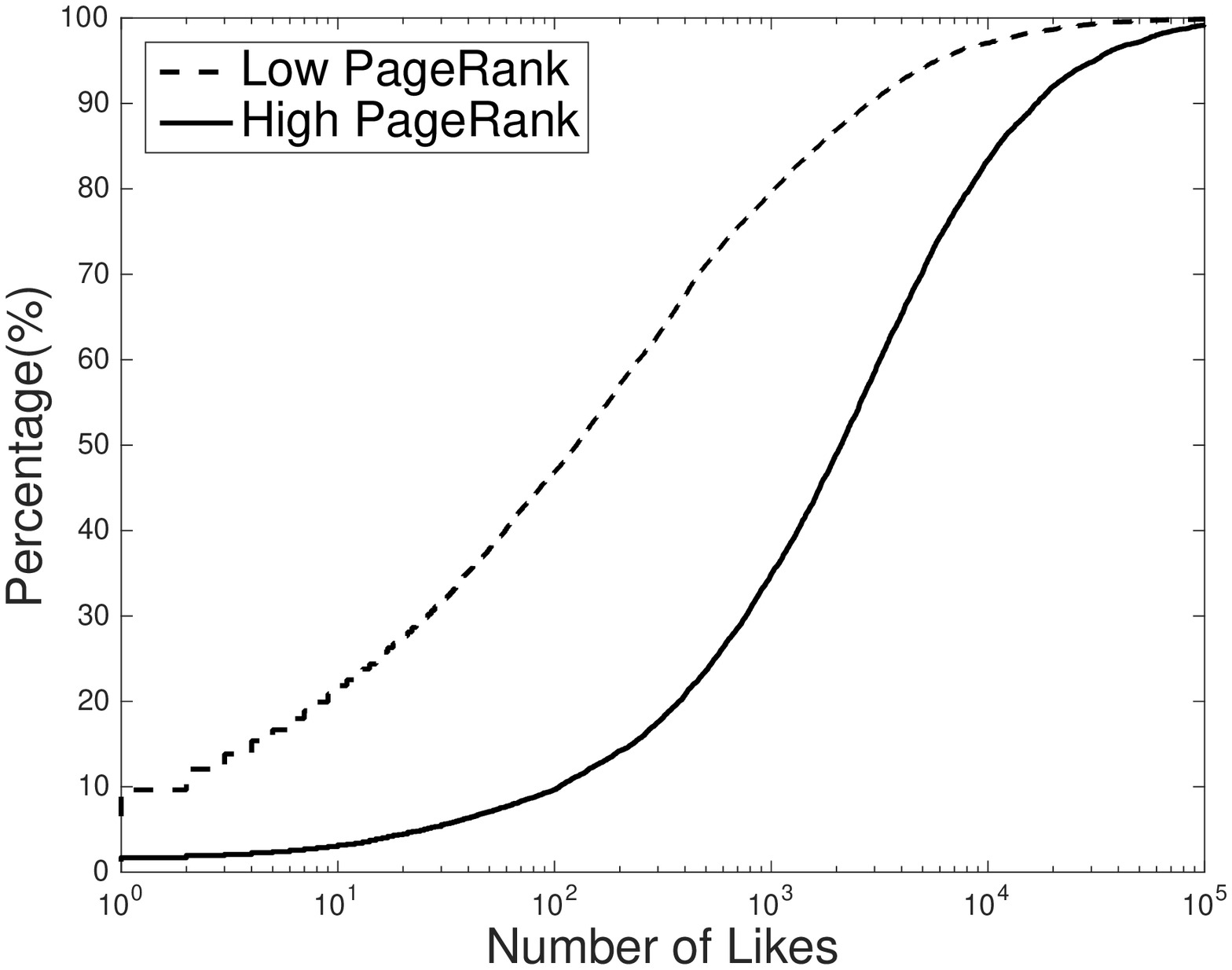}
\label{t_likes}}

\subfloat[Lists]{\includegraphics[width=.45\linewidth]{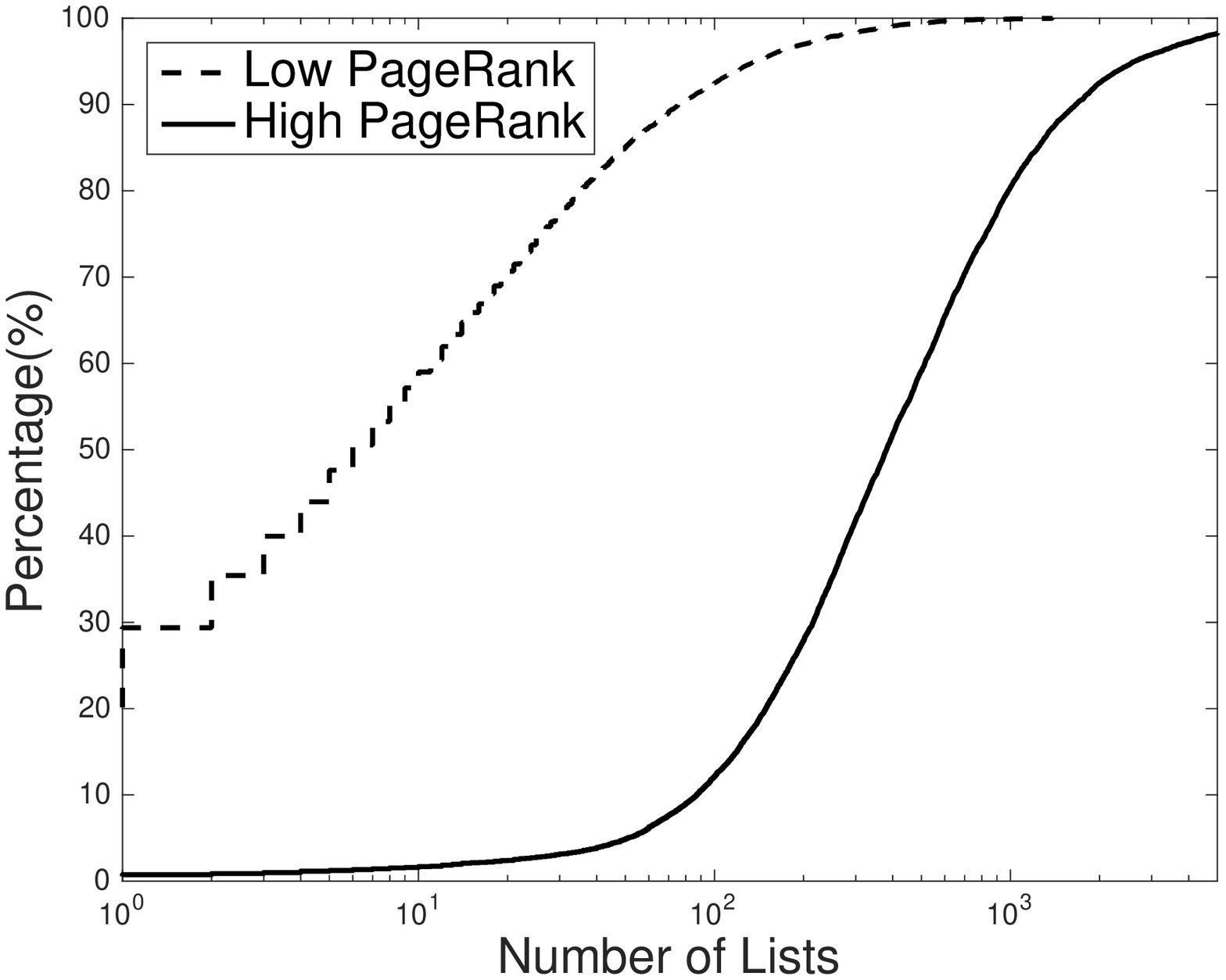}
\label{t_lists}}
\hfill
\subfloat[Bio]{\includegraphics[width=.45\linewidth]{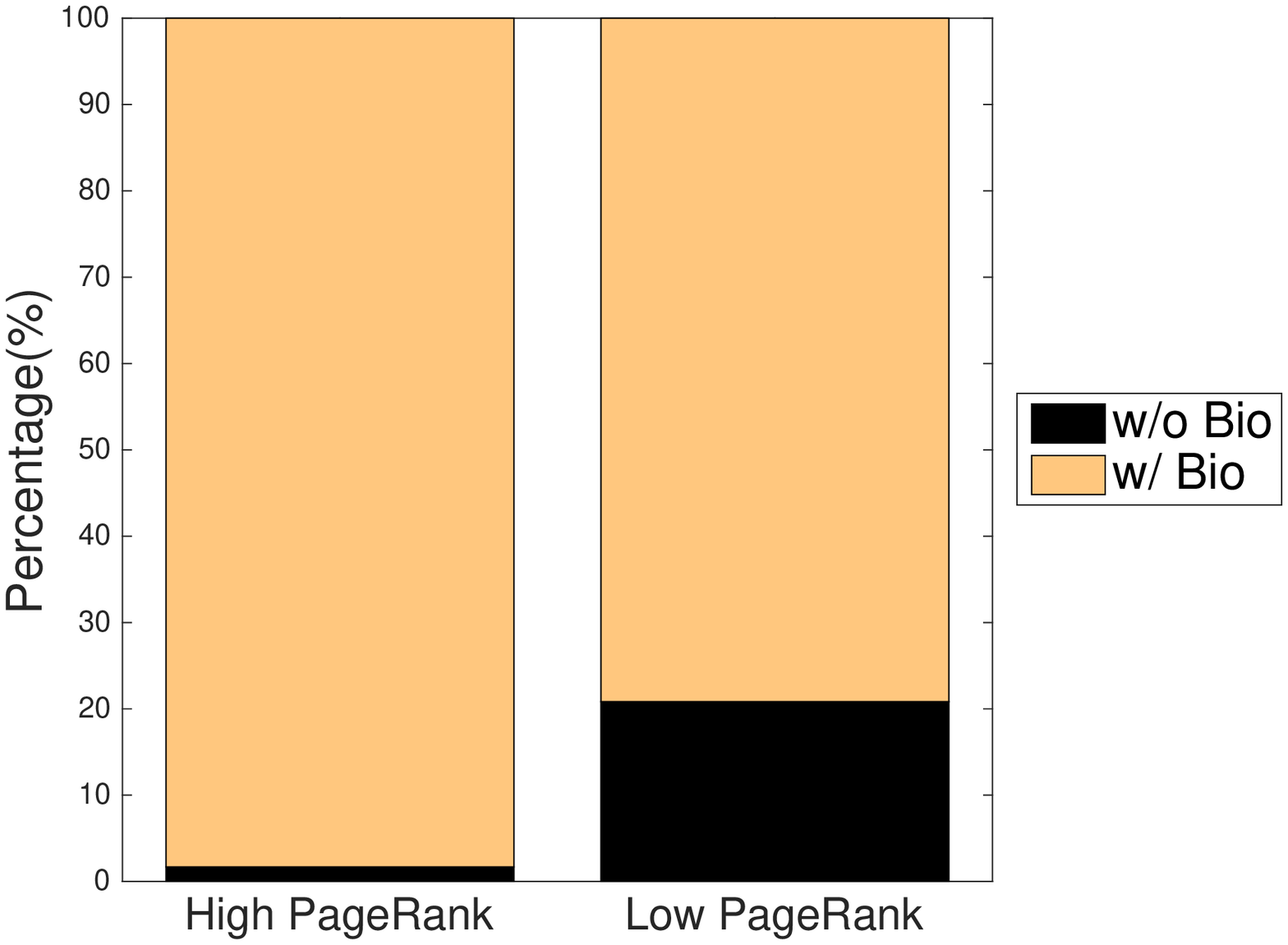}
\label{t_bio}}

\caption{Comparison between High and Low PageRank Medium Users based on Selected Features}
\label{t_comparison}
\end{figure}

\subsection{Prediction and Attraction of High PageRank Users}
\label{prediction_subsection}

Even if a new OSN has a significant growth in its size, it would become a ``ghost town'' like Google+~\cite{Gonzalez:2013:GGD:2488388.2488431} if its users are not active enough. Thus it is of vital importance for new OSNs to attract potential high PageRank users as many as possible to enliven their social communities. To achieve this purpose, we build a machine-learning-based classification model to predict high PageRank users in new OSNs using only data of other well-established OSNs. We also design a system based on this model for new OSNs to identify and attract potential high PageRank users from other well-established OSNs through the cross-site linking function.

Our classification model is designed to use a user's data on one OSN to predict her PageRank on the other OSN that the user would join through the cross-site linking function in the future.
In Medium's case, we intend to predict high PageRank Medium users defined in Section~\ref{csl_subsection} based on their Twitter or Facebook data only. Due to the privacy policy and setting of Facebook, it is hard to collect user data from it. Thus we use Twitter data described in Section~\ref{data_collection_section} to predict the high PageRank Medium users.

Since only the ``TW only'' and ``Both'' users have linked to Twitter accounts, our prediction is restricted to these users. We train a classification model by supervised machine learning algorithms to predict high PageRank Medium users based on their Twitter data only. We first select six key features from Twitter data in our data set, as listed in Table~\ref{features_table}.

Before training, we compare high PageRank users and low PageRank users in terms of these features in Fig.~\ref{t_comparison}. We find significant differences between these two user groups, indicating that there is a great possibility that a high PageRank Medium user can be identified by these features.

We randomly pick 8,000 high PageRank users and 8,000 low PageRank users to form a training data set, and another 2,000 high PageRank users and 2,000 low PageRank users to form a testing data set. We train different models using the training data set, then evaluate them using the testing data set.

We use numerous machine learning algorithms to train classification models, including XGBoost~\cite{Chen:2016:XST:2939672.2939785}, Random Forest, C4.5 Decision Tree, Bayes Net, Naive Bayes, Sequential Minimal Optimization (SMO)~\cite{sequential-minimal-optimization-a-fast-algorithm-for-training-support-vector-machines} and Logistic Regression. Besides XGBoost, other algorithms are applied in Weka~\cite{Hall:2009:WDM:1656274.1656278}.

We apply the following four classical metrics, i.e., precision, recall, F1-score and the area under the curve (AUC), to evaluate the performance of classification models. Precision is the fraction of predicted high PageRank users who are really high PageRank users. Recall is the fraction of high PageRank users who are accurately predicted. F1-score is defined by the harmonic mean of precision and recall, i.e., $F_1 = 2 \cdot \frac{precision \cdot recall}{precision + recall}$. AUC is the area under the receiver operating characteristic (ROC) curve~\footnote{The receiver operating characteristic (ROC) curve is created by plotting the true positive rate (TPR) against the false positive rate (FPR) at various threshold settings, which illustrates the diagnostic ability of a binary classifier system as its discrimination threshold is varied.}, which equals the probability that a classifier will rank a randomly chosen positive instance higher than a randomly chosen negative one.~\footnote{https://en.wikipedia.org/wiki/Receiver\_operating\_characteristic}

For each set of parameters during training, we use 10-fold cross-validation~\footnote{In 10-fold cross-validation, the training and validation data set is randomly divided into 10 subsets with equal size. Of the 10 subsets, a single subset is retained as the validation data for evaluating the model, and the remaining 9 subsets are used for training. The cross-validation process is repeated 10 times, with each of the 10 subsets used once as the validation data.} to calculate the four metrics. For each algorithm, we carefully tune the parameters and record the ``best'' parameters which could achieve the highest F1-score. Then we use the ``best'' parameters to predict high PageRank users using the testing data set.

\begin{table*}[!t]
\renewcommand{\arraystretch}{1.3}
\caption{Prediction of High PageRank Medium Users}
\label{prediction_table}
\centering
\begin{tabular}{c|c|c c c c}
\hline
Algorithm & Parameter & Precision & Recall & F1-score & AUC\\
\hline
                    & learning\_rate=0.37, max\_depth=6, min\_child\_weight=1, gamma=0, &       &       &       &       \\
XGBoost             & subsample=0.6, colsample\_bytree=0.9, alpha=0.005, lambda=1,      & 0.939 & 0.944 & 0.942 & 0.986 \\
                    & booster=gbtree, objective=multi:softmax/softprob, num\_class=2             &       &       &       &       \\
Random Forest       & 100 trees, features/tree -K=3, max depth -depth=9                 & 0.933 & 0.936 & 0.934 & 0.980 \\
C4.5 Decision Tree  & confidence factor -C=0.18, min instance/leaf -M=9                 & 0.927 & 0.925 & 0.926 & 0.959 \\
Bayes Net           & search algorithm -Q=EMBC                                          & 0.909 & 0.936 & 0.922 & 0.968 \\
Naive Bayes         & default                                                           & 0.893 & 0.929 & 0.910 & 0.956 \\
SMO                 & kernel -K=PolyKernel, complexity -C=975                           & 0.940 & 0.868 & 0.903 & 0.906 \\
Logistic Regression & number of boosting iterations -I=81                               & 0.944 & 0.848 & 0.893 & 0.962 \\
\hline
\end{tabular}
\end{table*}

\begin{table}[!t]
\renewcommand{\arraystretch}{1.3}
\caption{Selected Features}
\label{features_table}
\centering
\begin{tabular}{c|c}
\hline
Feature & Description\\
\hline
Followings & the number of followings for a user in Twitter\\
Followers  & the number of followers for a user in Twitter\\
Tweets     & the number of tweets for a user in Twitter\\
Likes      & the number of likes for a user in Twitter\\
Lists      & the number of lists for a user in Twitter\\
Bio        & 1 if a user enables Twitter biography, 0 otherwise\\
\hline
\end{tabular}
\end{table}

\begin{table}[!t]
\renewcommand{\arraystretch}{1.3}
\caption{$\chi^2$ Statistic}
\label{chi_table}
\centering
\begin{tabular}{c|c|c}
\hline
Rank & Feature & $\chi^2$ \\
\hline
1 & Lists      & 11484.35 \\
2 & Followers  & 11461.27 \\
3 & Tweets     & 5395.73  \\
4 & Likes      & 4314.13  \\
5 & Followings & 3735.51  \\
6 & Bio        & 1469.19  \\
\hline
\end{tabular}
\end{table}

Table~\ref{prediction_table} shows the performance of each algorithm on the testing data set. We can see that XGBoost performs the best, which achieves an F1-score of 0.942 and an AUC of 0.986. Thus, we conclude that our model can predict high PageRank Medium users based on Twitter data only. We notice that even the simple Logistic Regression has a good performance, which indicates that our feature selection is quite successful.

To evaluate the discriminative power of selected features, we present their rank by $\chi^2$ (Chi Square) statistic~\cite{Yang:1997:CSF:645526.657137} in Table~\ref{chi_table}. We can see that the most powerful features are the number of lists, followers and tweets. The number of followers can partly reflect a user's influence in Twitter. The number of lists and tweets can illustrate a user's activity in Twitter. In short, users tend to perform similarly in Twitter and Medium.

\begin{figure*}[!t]
\centering

\includegraphics[width=\linewidth]{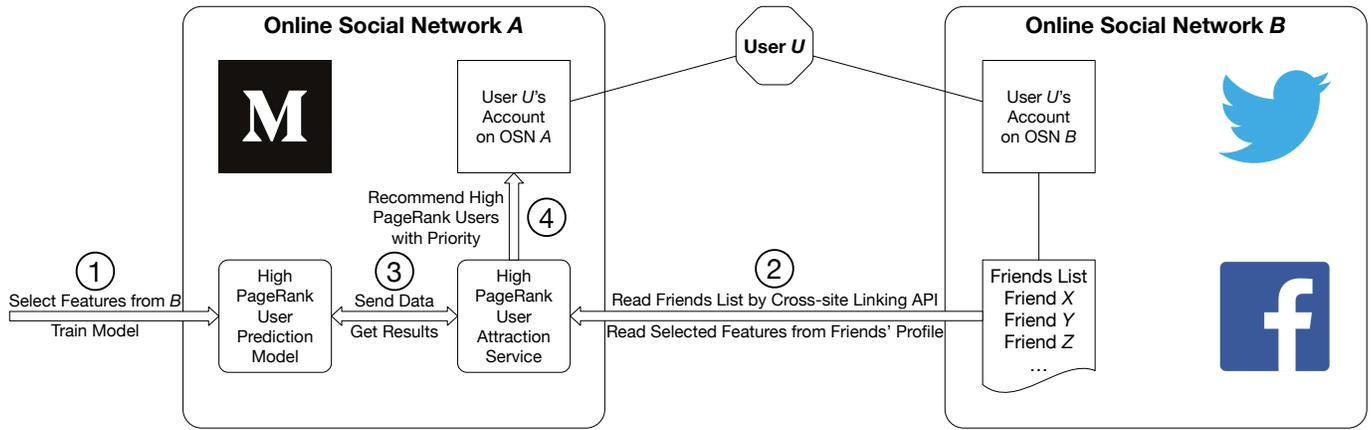}
\caption{Design of the ``High PageRank User Attraction System''}
\label{system}

\end{figure*}

To assist new OSNs to attract more high PageRank users from other well-established OSNs through the cross-site linking function, we design the ``High PageRank User Attraction System'' based on our prediction model. In Fig.~\ref{system}, $A$ is a new OSN like Medium and $B$ is an well-established OSN like Twitter and Facebook. $U$ is a user who has linked her account on $A$ to her account on $B$ through the cross-site linking function. $A$ launches ``High PageRank User Attraction Service'' to attract high PageRank users from $B$ by the following steps. First, it selects features from $B$ and trains the ``High PageRank User Prediction Model''. Second, it reads $U$'s friends list on $B$ and collects selected features from $U$'s friends' profiles on $B$. Third, it sends the collected features to ``High PageRank User Prediction Model'' and gets the prediction results of high PageRank users. Fourth, it places these users in priority when it recommends $U$ to invite her friends from $B$ to $A$. Thus, the system helps $A$ to purposely attract high PageRank users from $B$ to enliven its social community.

%% file: conclusion.tex
\section{Conclusion and Future Work}
\label{conclude_section}

This paper provides a comprehensive study of the service integration of online social networks (OSNs). To the best of our knowledge, this is the first work to analyze Medium's social graph. We find that the service integration can change the graph structure of the new OSN, and bring new users from other well-established OSNs to the new one through the cross-site linking function. However, we find that almost none of the new users would become high PageRank users in the new OSN, which is a shortcoming of the service integration. To solve this problem, we build a machine-learning-based model to predict high PageRank users. It achieves a high F1-score of 0.942 and a high AUC of 0.986. Then we extend this model to a system to assist new OSNs to attract highly PageRank users from other well-established OSNs through the cross-site linking function.

However, we only examine our prediction model on Twitter and Medium, which is just a proof of concept that the high PageRank users can be predicted. In the future, we will apply our prediction methods to more OSNs to verify that the prediction model can be extended to other sites.
This work also shows that we are able to connect user identities on multiple OSNs to the same person, which allows us to study the differences and similarities of the same user across multiple OSNs. We will investigate these differences and similarities across Medium, Twitter and Facebook in the future work.